\documentclass[fleqn,usenatbib,usedcolumn]{mnras}

\usepackage{newtxtext,newtxmath}

\usepackage[T1]{fontenc}

\DeclareRobustCommand{\VAN}[3]{#2}
\let\VANthebibliography\thebibliography
\def\thebibliography{\DeclareRobustCommand{\VAN}[3]{##3}\VANthebibliography}

\usepackage{graphicx}	
\usepackage{amsmath}	

\usepackage[british]{babel}             
\usepackage{newtxtext}                  
\usepackage[T1]{fontenc}                
\usepackage{amsfonts}
\usepackage{amsbsy}
\usepackage{bm}
\usepackage{url}
\usepackage{microtype}
\usepackage{rotating}
\usepackage{sidecap}
\usepackage{longtable}
\usepackage{lscape}
\usepackage[export]{adjustbox}
\usepackage{xcolor}
\usepackage{colortbl}
\usepackage{booktabs}
\usepackage[flushleft]{threeparttable}

\let\la=\lesssim 

\def\cm2{{\rm cm^{-2}}}

\title[MeerKAT caught a Mini Mouse]{MeerKAT caught a Mini Mouse: serendipitous detection of a young radio pulsar escaping its birth site} 

\author[S. Motta et al.]{Motta, S.~E.$^{1, 2}$, Turner, J. D.$^{3}$, Stappers, B.$^{3}$, Fender, R. P.$^{2}$, Heywood, I.$^{2,4,5}$, Kramer, M.$^{6}$, Barr, E.D.$^{6}$\\
$^1$Istituto Nazionale di Astrofisica, Osservatorio Astronomico di Brera, via E.\,Bianchi 46, 23807 Merate (LC), Italy\\
$^2$University of Oxford, Department of Physics, Astrophysics, Denys Wilkinson Building, Keble Road, OX1 3RH, Oxford, United Kingdom\\
$^3$Jodrell Bank Centre for Astrophysics, Department of Physics and Astronomy, The University of Manchester, Manchester M13 9PL, UK\\
$^{4}$Department of Physics and Electronics, Rhodes University, PO Box 94, Makhanda 6140, South Africa\\
$^{5}$South African Radio Astronomy Observatory, Cape Town, South Africa. \\
$^{6}$ Max-Planck-Institut fur Radioastronomie, Auf dem Hügel 69, D-53121, Bonn, Germany\\
}

\date{Accepted XXX. Received YYY; in original form ZZZ}

\pubyear{2020}

\begin{document}
\label{firstpage}
\pagerange{\pageref{firstpage}--\pageref{lastpage}}
\maketitle

\begin{abstract}

\noindent In MeerKAT observations pointed at a Galactic X-ray binary located on the Galactic plane we serendipitously discovered a radio nebula with cometary-like morphology. The feature,  which we named `the Mini Mouse' based on its similarity with the previously discovered `Mouse' nebula, points back towards the previously unidentified candidate supernova remnant G45.24$+$0.18. We observed the location of the Mini Mouse with MeerKAT in two different observations, and we localised with arcsecond precision the 138 ms radio pulsar PSR J1914+1054g, recently discovered by the FAST telescope, to a position consistent with the head of the nebula.
We confirm a dispersion measure of about 418 pc cm$^{-3}$ corresponding to a distance between 7.8 and 8.8 kpc based on models of the electron distribution. Using our accurate localisation and 2 period measurements spaced 90 days apart we calculate a period derivative of (2.7 $\pm$ 0.3) $\times$ 10 $^{-14}$ s s$^{-1}$. We derive a characteristic age of approximately 82 kyr and a spin down luminosity of 4~$\times$~10$^{35}$ erg s$^{-1}$, respectively. 
For a pulsar age comparable with the characteristic age, we find that the projected velocity of the neutron star is between 320 and 360 km/s if it was born at the location of the supernova remnant. The size of the proposed remnant appears small if compared with the pulsar characteristic age, however the relatively high density of the environment near the Galactic plane could explain a suppressed expansion rate and thus a smaller remnant.

\end{abstract}



\section{Introduction}

Pulsar wind nebulae (PWN) are the result of the interaction between the relativistic particle winds produced by rotationally-powered pulsars and the surrounding medium.  
The rapid rotation of the magnetic field of the neutron star (NS) powers a relativistic wind, which, via interaction with the ambient medium, generates a termination shock downstream of which synchrotron radiation emerges with a spectrum extending from the radio band up to the $\gamma$-rays \citep[see, e.g.,][for a review]{Gaensler2006}. Sometimes a PWN is surrounded by a shell-like supernova remnant (SNR), legacy of the explosion that gave birth to the pulsar itself, and in this case the system is termed `composite' \citep{Matheson2005}.

If a pulsar moves through the interstellar medium (ISM) at a supersonic speed, a bow shock forms, which redirects and channels the pulsar wind in the direction opposite to that of the pulsar’s motion. This may result in a detectable  tail which can extend for several parsecs behind the NS \citep[see, e.g.,][]{Kargaltsev2015}. In such cases, the SNR associated with the formation of the NS may be located parsecs away from the PWN, although the cometary tail may either connect to or point in the direction of the supernova shell. Among the several tens of PWN known, only a small number of objects present such a cometary morphology, which is indicative of high proper velocities and/or a dense ISM. High spatial resolution observations, in radio and in the X-ray band are adding to this group \citep[e.g.,][]{Klingler2018}. 

Detailed studies of the PWNe and their pulsars can provide crucial information on these systems, as the PWNe's appearance, spectrum, and radiative efficiency depend on the pulsars’ parameters (e.g., spin, spin-down power, surface magnetic field and its orientation), on the pulsars’
velocity, and on the properties of the pulsar wind (e.g., flow speed and magnetisation). 
Additionally, since PWNe have a well-defined central engine and are often close enough to be resolvable with high angular resolution observations, they represent excellent laboratories for studying both relativistic particle winds and the shocks that result when such outflows collide with the ISM, thus offering the opportunity to constrain the properties of their environment.

\bigskip

As part of the ThunderKAT Large Survey Program, which is aimed at providing a long-term view of interesting transients in the radio band, we observed the field of the black hole binary GRS 1915+105 with the MeerKAT radio telescope at L-band with a few arcsec angular resolution \citep{Motta2021}.
In the MeerKAT field of the source we identified a feature that closely resembles `the Mouse' \citep{Yusef-Zadeh1987}, a radio nebula with axial symmetry, consisting of a bright `head' and a long `tail', observed in the direction of the Galactic Centre region, which hosts the young radio pulsar PSR J1747$-$2958 \citep{Camilo2002}. Based on the resemblance with the Mouse, we named the newly identified feature in the GRS 1915+105 field `the Mini Mouse'. 

This paper is structured as follows; in \S \ref{sec:Obs} we describe the imaging and time-domain data acquisition. In \S \ref{sec:results} we report the discovery of the Mini Mouse, the identification of PSR J1914+1054g, recently discovered by FAST, as the counterpart, and calculate the spin period derivative. Finally, in \S \ref{sec:discussion}, we discuss the inferred nature of the Mini Mouse from these measurements.

\section{Observations and data Analysis}\label{sec:Obs}

\subsection{Continuum observations}

We observed the field of GRS~1915+105  61 times with MeerKAT as part of the ThunderKAT Large Survey Project \citep{Fender2016b} between December 2018 and April 2022. We observed at a central frequency of 1.28 GHz across a 0.86 GHz bandwidth (856 – 1712 MHz). The correlator delivered either 4096 or 32768 channels, with a 8-s integration time per visibility point, which were binned down to 1024 channels for consistency before any further analysis. Between 58 and 64 of the 64 available dishes were used in the observations, with a maximum baseline of 7.698 km. 
Of 61 observations, 60 consisted of 15 minutes of on-source time, book-ended by two 2~min scans of the secondary calibrator J2011$-$0644, plus a 10~min observation of a primary  calibrator (J1939$-$6342) \citep{Motta2021}. One observation lasted 90~minutes, of which 60 min was on-source, 20 minutes on the primary calibrator, and 3 minutes on the secondary calibrator. 

Imaging was conducted via a set of \textsc{Python} scripts specifically tailored for the semi-automatic processing of MeerKAT data (\textsc{OxKAT}\footnote{\url{https://github.com/IanHeywood/oxkat}}, \citealt{Heywood2020}). Initial flagging to remove the first and final 100 channels from the observing band, autocorrelations, zero amplitude visibilities, and RFI was performed in \textsc{CASA} \citep{McMullin2007}. Further flagging was performed using the \textsc{TRICOLOUR} package \footnote{\url{https://github.com/ska- sa/tricolour/}}, after averaging the data in time (8~s) and frequency (8 channels) for imaging purposes.
We imaged the field of GRS 1915+105 using \textsc{WSClean} (\citealt{Offringa2012}) and combining the visibilities from all our observations after uv-subtracting the variable emission from GRS 1915+105. Direction independent self-calibration was performed using \textsc{CUBICAL} \citep{Kenyon2018} by solving for phase and delay corrections for every 32 seconds of data.

\subsection{Search for a pulsar counterpart}

The most sensitive pulsar search previously conducted that covered the position of the Mini Mouse is the ongoing FAST Galactic Plane Pulsar Snapshot (GPPS) survey \citep{Han2021}, with a 10 $\upsigma$ sensitivity threshold between 1 and 10 $\upmu$Jy for periods of 10$^{-2}$ and 10~s. 
The FAST GPPS survey recently discovered a faint ($S_{1400}$ = 33~$\upmu$Jy) pulsar, PSR J1914+1054g (hereafter J1914) with a spin period of 138~ms, and a position consistent with the head of the Mini Mouse within a 1.5\arcmin~uncertainty \citep{Han2021}.

In order to confirm the association between either J1914 or an undiscovered pulsar and the Mini Mouse head, we conducted a 2 hr observation (henceforth referred to as OBS1) as part of the TRAPUM Large Survey project \citep{Stappers2016} with a nominal 10 $\upsigma$ sensitivity of 10.6 $\upmu$Jy. We observed at a central frequency of 1284 MHz and a bandwidth of 856 MHz (identical to that used in the continuum observations), split into 4096 channels, with a sampling time of 153 $\times$ 10$^{-6}$~s.
We used all 64 available dishes which gives the minimum coherent beam (CB) size. A CB is formed by coherently summing the digitised signals from each receiver after correcting for the geometric delay, and accesses a much smaller field of view than the primary (incoherent) beam does. The on-site Filterbank and BeamForming User-Supplied Equipment (FBFUSE) computer cluster, developed by the Max-Planck Instit{\"u}t f{\"u}r Radioastronomie \citep{Barr2017}, uses the beamforming package \textsc{mosaic}\footnote{\url{https://github.com/wchenastro/Mosaic} by Weiwei Chen} to derive a set of delay polynomials that will synthesise multiple CBs. This enables the tessellation of hexagonally packed beams across a source \citep{Chen2021}. For OBS1, 266 beams were arranged in a rectangle overlapping at their 75 per cent power width, to densely cover the full length and breadth of the nebula. The beam at the centre of the field of view was placed on the head of the Mini Mouse at the position of largest continuum flux. An additional 66 beams were used to tile over the portion of the 1.5\arcmin~position error region that was not covered by the rectangle.

Approximately 45 minutes into the observation, a failure of one of the 8 capture nodes of the Accelerated Pulsar Search User Supplied Equipment (APSUSE) resulted in ensuing data being corrupted in some beams. The observation was suspended a short time later. Luckily, the 7 most central beams positioned on the head of the Mini Mouse were unaffected, yielding 104 minutes of good data. The corresponding filterbanks were processed offline 
to search for J1914 directly before any global periodicity search was attempted.

Raw data is recorded on the APSUSE file system in the \texttt{SIGPROC}\footnote{\url{http://sigproc.sourceforge.net/}} filterbank format, segmented by beam and in 5 minute chunks. The initial search for J1914 was conducted first on the innermost beam as it is located both at the point of maximum sensitivity (boresight) and at the position of maximum flux of the nebula's head. The \texttt{filtool} program, which is part of \textsc{pulsar}X\footnote{\url{https://github.com/ypmen/PulsarX} by Yunpeng Men}, was used to clean Radio Frequency Interference (RFI) and to stitch together the continuous data into a single filterbank file. Using \textsc{psrchive}'s \textsc{dspsr} \citep{VanStraten2011}, the filterbank file was folded using an ephemeris containing J1914's period and the DM for the epoch reported in \cite{Han2021}. \textsc{clfd}\footnote{\url{https://github.com/v-morello/clfd} by Vincent Morello} was then applied to the full resolution folded data to clean any lingering RFI. Finally, a search over period and DM was performed with \texttt{pdmp} \citep{VanStraten2011, Hotan2004} with a window of 50 $\upmu$s in period and 20 units of DM. 
The observation yielded a detection of J1914 with a significance of 19.4 in the beam directly placed on the Mini Mouse head, with $P$ = 138.867651~ms and a DM of 418.0 $\pm$ 0.3 pc cm$^{-3}$. As a result of this detection at the head of the nebula, no full scale periodicity search of all CBs was attempted. 
Using the spin period measured by the FAST telescope about 2 years before\footnote{The period measured by FAST was kindly provided by Prof. JinLin Han via private communication.} 
we estimated a tentative period derivative of about 3~$\times$~10$^{-14}$~s~s$^{-1}$. 

In order to obtain a second accurate measurement of the spin period of J1914 with MeerKAT, and to meaningfully constrain the spin derivative, we requested and obtained a Director's Discretionary Time (DDT) observation of the Mini Mouse position (henceforth OBS2, project code DDT-20221028-
SM-01). Only 7 beams at 75 per cent overlap were required to cover J1914's position derived from OBS1. The optimal tiling pattern selected by \textsc{mosaic} is a central beam surrounded by a hexagon of 6, a configuration suited for signal-to-noise-based localisation techniques. 
The reduction in the number of beams also allowed for observing at a time resolution of 76$\times$10$^{-6}$~s. These data were reduced and searched in the same way as in OBS1. The observing parameters and resources used for both observations are summarised in Tab. \ref{tab:fits}.

\begin{table*}
\begin{center}
\caption{A summary of the observing parameters and resources used for the two MeerKAT observations of J1914. $^{\dagger}$ The value refers to the co-ordinates of the central CB.}\label{fig:obs}
\begin{tabular}{lll}

\hline
Parameter                               & OBS1         & OBS2 \\
\hline \hline
UTC Start Time (yyyy-MM-dd-hh:mm:ss)    & 2022-08-03-21:03:30       &   2022-11-01-16:54:47 \\
Start MJD                               & 59794.877431              &   59884.704630        \\
Duration (s)                            & 6296                      &   7186                \\
Number of dishes                            & 64                      &   62                \\
Central frequency (MHz)                 & 1284                      &   1284                \\
Bandwidth (MHz)                         & 856                       &   856                 \\
Channels                                & 4096                      &   4096                \\
Sampling time (s)                       & 156 $\times$ 10$^{-6}$     &   72 $\times$ 10$^{-6}$ \\
Right Ascension† (hh:mm:ss)              & 19:14:09.46               &   19:14:09.66         \\
Declination† (dd:mm:ss)                 & +10:54:43.7               &   +10:54:42.2         \\

\hline \hline
\end{tabular}
\end{center}
\end{table*}

\bigskip


\begin{table*}
\begin{center}
\caption{Measured and inferred quantities for J1914 derived from OBS2. The value in brackets denotes the 1 $\upsigma$ uncertainty on the rightmost digit.  
}\label{tab:fits}
\begin{tabular}{ll}
\hline\hline

\multicolumn{2}{c}{Measured quantities} \\ 
\hline
Right Ascension, $\alpha$ (J2000) \dotfill \dotfill        &   19$^\text{h}$14$^\text{m}$09\fs83$^{+0.41}_{-0.12}$ \\ 
Declination, $\delta$ (J2000) \dotfill                  & +10\degr54\arcmin35.9\farcs$^{+13.4}_{-2.0}$ \\ 
Pulse period, $P$ (s) \dotfill                          & 0.138867860(17)\\
Pulse period derivative $\dot{P}$ (s s$^{-1}$) \dotfill         & 2.7(3) $\times$10$^{-14}$\\
Duty cycle (\%) \dotfill  & 14\\ 
Dispersion measure, DM (pc cm$^{-3}$) \dotfill          &  418.90(26)\\
Flux density, $S_{\mathrm{1284}}$ ($\upmu$Jy) \dotfill           &  62.1   \\

\hline
\multicolumn{2}{c}{Inferred quantities} \\ 
\hline
Distance (\textsc{NE2001}) (kpc), $d_{1}$ \dotfill        		  & 7.8\\
Distance (\textsc{YMW16}) (kpc), $d_{2}$ \dotfill        		  & 8.8\\
Characteristic age, $\tau_c$ (kyr) \dotfill                          			  & 82(11)\\
Surface dipole magnetic field strength, $B$ (G)  \dotfill    & $1.9(1)\times10^{12}$\\
Spin-down luminosity, $\dot{E}$~(erg~$\mathrm{s}^{-1}$) \dotfill                   & $4.0(6)\times10^{35}$\\
Radio pseudo-luminosity, L$_{\rm 1284}$ at $d_{1}$ ($m\mathrm{Jy\,kpc^{2}}$) \dotfill & 3.8\\
X-ray Luminosity, L$_{\rm X} (1-10 \, \rm {keV})$ at $d_{1}$ (erg s$^{-1}$) \dotfill  &  $< 7~\times$~10$^{32}$\\

\hline \hline
\end{tabular}
\end{center}
\end{table*}

\begin{figure*}
    \centering
    \includegraphics[width=1.0\textwidth]{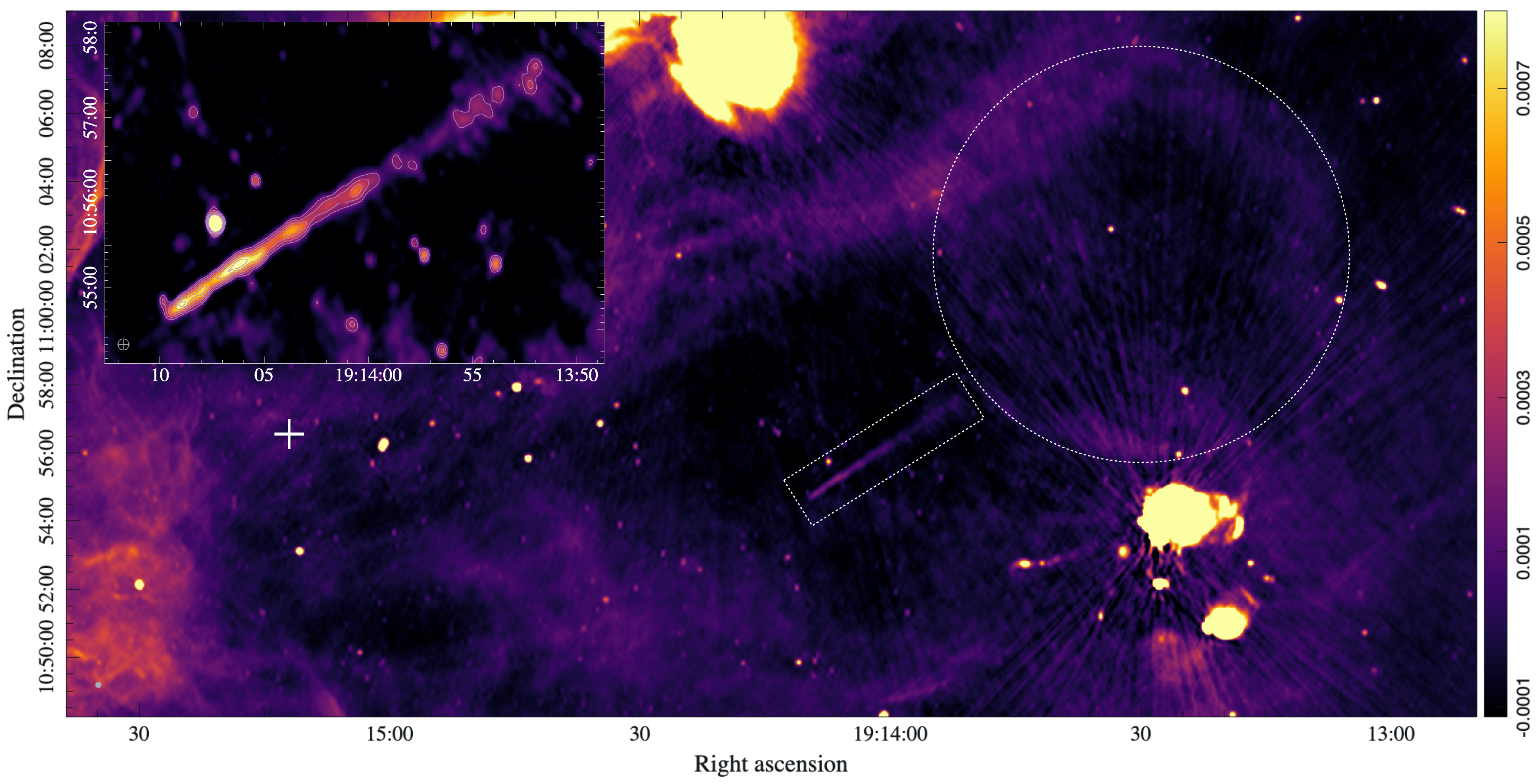}
    \caption{A portion of the field centred on GRS 1915+105 as seen by the MeerKAT radio telescope at 1.28 GHz. The total on-source time is 15.5 hr. The restored beam for this map is circular and has a radius of 6.6", shown in the bottom left corner of the image. The Mini Mouse and the SNR associated with the birth of pulsar J1914 are marked by a rectangle and circle, respectively. The location of GRS 1915+105 is marked by a cross. Inset: a zoom into the Mini Mouse. Contours are equally spaced, and span the range 10--200 uJy. The `tooth' at the edge of the nebula is likely a background source. }
    \label{fig:cont}
\end{figure*}
\begin{figure}
    \centering
    \includegraphics[width=1.0\columnwidth]{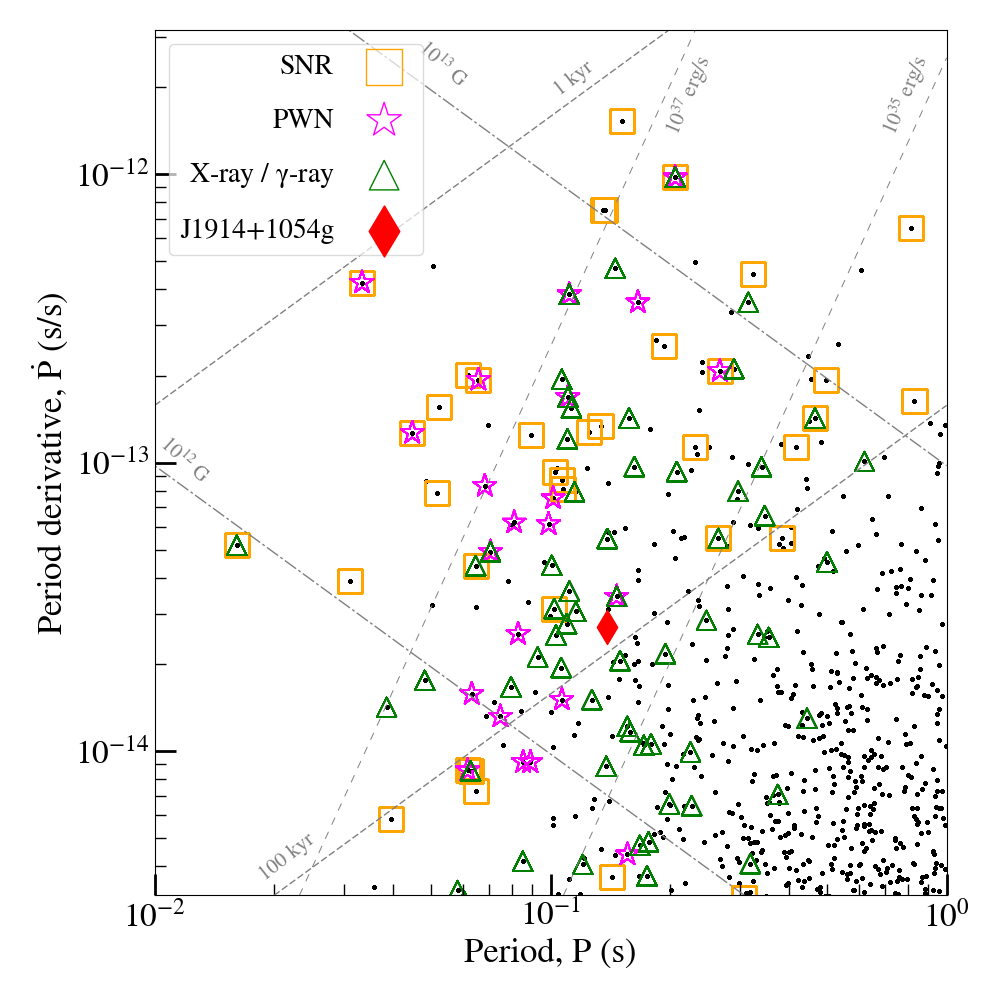}
    \caption{A plot of $\dot{P}$ against $P$ for the young pulsar populace, J1914 inclusive. The canonical radio pulsars emerge towards the lower right of the plot. Pulsars with confirmed associations with SNRs, PWNe or detections at high energies, according to the ATNF Catalogue \citep{Manchester2005} are indicated, as are the contour lines for $B$, $\tau_{\text{c}}$ and $\dot{E}$.}
    \label{fig:PPdot}
\end{figure}

\begin{figure}
    \centering
    \includegraphics[width=1.0\columnwidth]{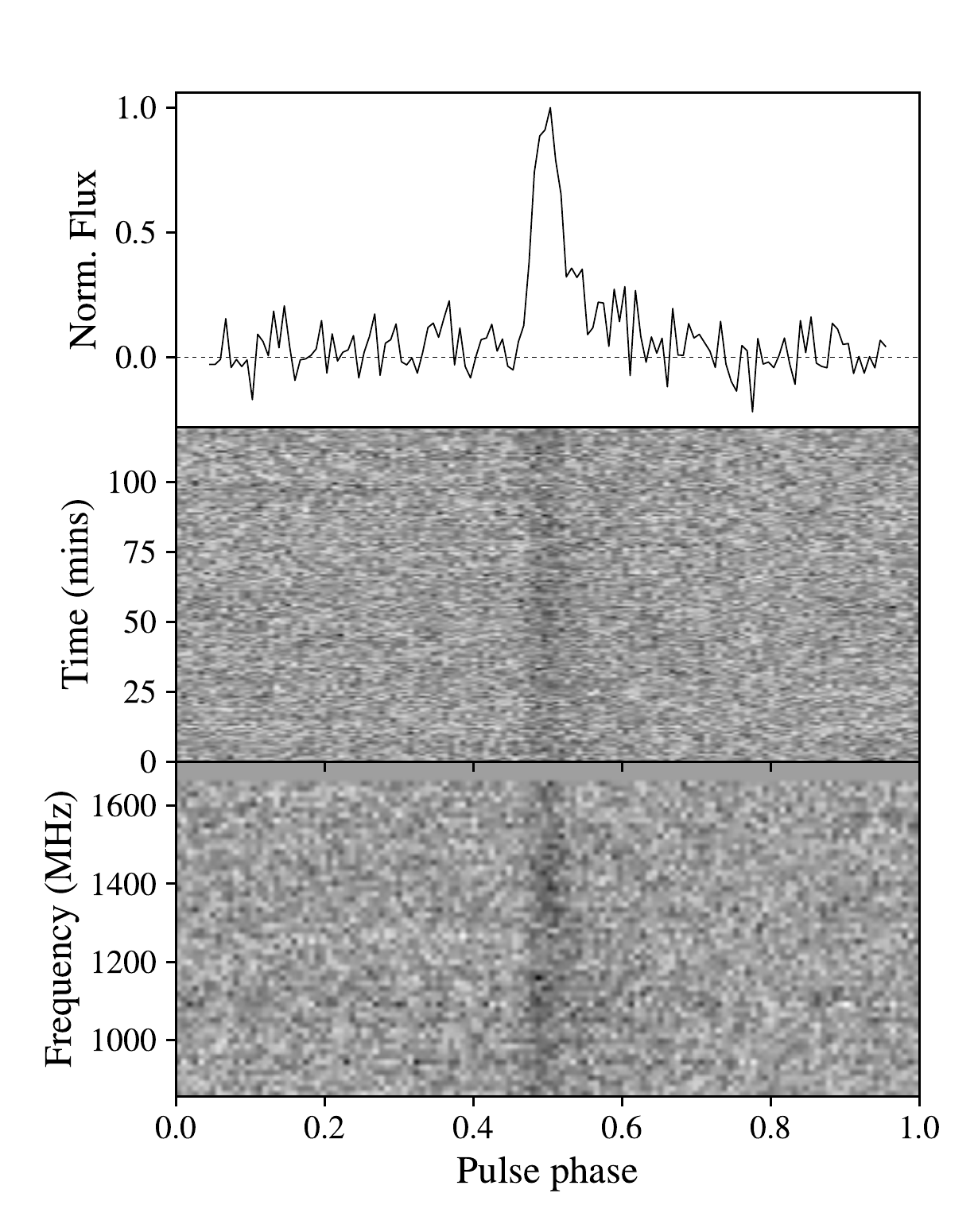}
    \caption{Second MeerKAT detection of J1914 folded at the best period and dedispersed.  Bottom panel: Frequency against pulse phase across the band with a resolution of 128 channels. Middle panel: Band-integrated time series for the 2 hr observation showing the flux for each 2 minute sub-integration. Top panel: Integrated pulse profile made by stacking the time series into 128 bins. Evidence for scattering in the ISM can be seen in the asymmetric broadening of the pulse profile.}
    \label{fig:pulse}
\end{figure}

\begin{figure*}
    \centering
    \includegraphics[width=\columnwidth]{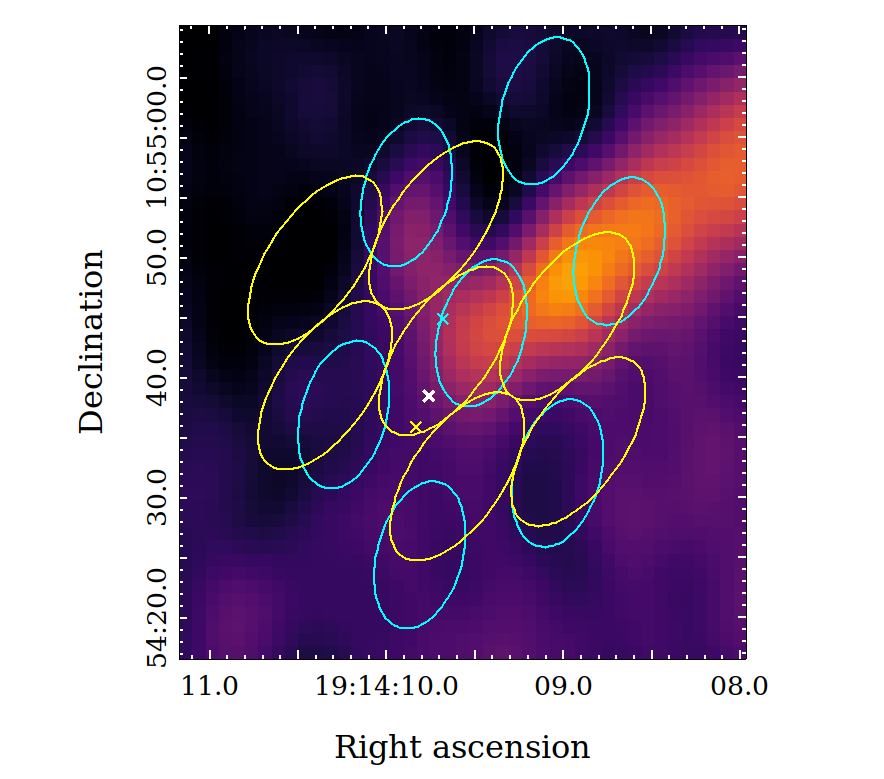}
    \includegraphics[width=\columnwidth]{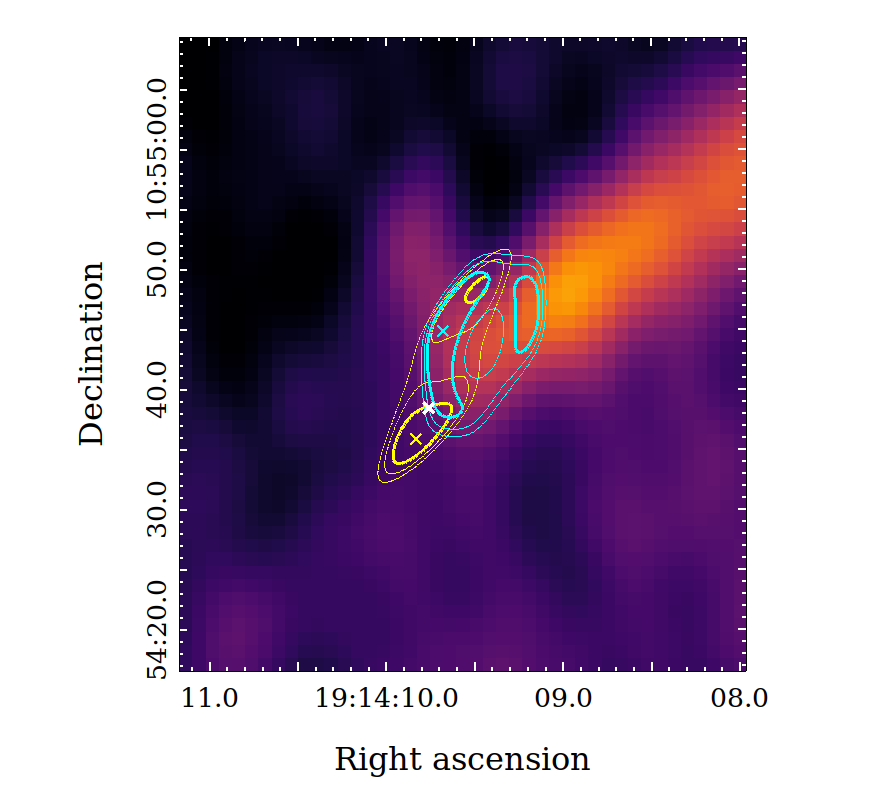}
    \caption{The OBS1 (cyan) and OBS2 (yellow) localisation of J1914 from S\textsc{ee}KAT. The crosses indicate the maximum likelihood position. The bold white cross is the weighted mean of the two positions. The ellipses (left) are the 75 per cent power radii of the CBs. The contours (right) are the 1 (bold), 2 and 3 $\upsigma$ levels of the likelihood fit.}
    \label{fig:localisation}
\end{figure*}

\section{Results}\label{sec:results}

An image of the MeerKAT field is shown in Fig. \ref{fig:cont}, which also includes a zoom into the Mini Mouse. 
The Mini Mouse appears as a long structure extending from position RA, DEC $19^{\text{h}}14^{\text{m}}09.5^{\text{s}}$, +10\degr54\arcmin42.7\arcsec~for over 5\arcmin~in the East-North-East direction, with a flux density varying along its axis between about 0.08 and 0.2~mJy~beam$^{-1}$ (marked by a rectangle in Fig. \ref{fig:cont}). Unlike the original Mouse, the Mini Mouse does not have an obviously extended head at the MeerKAT resolution of 6.6\arcsec), and remains unresolved in the transverse direction for its entire length, albeit brighter in the first 2\arcmin~than in the tail. Further observations at higher angular resolution and/or higher frequency will help to clarify the finer angular scale structure of this nebula. 

Our continuum radio map shows several bright extended structures, including HII regions and several shell-like non-thermal radio structures, many of which can be associated with supernova remnants (SNRs). One such structure is located to the East-North-East of the Mini Mouse, and is a previously unidentified dim, remarkably circular SNR candidate G45.24$+$0.18 (marked by a circle in Fig. \ref{fig:cont}. The geometrical centre of the SNR is located within 30\arcsec~from the extension of the Mini Mouse axis of symmetry, and 12\arcmin~away from the head of the Mini Mouse. The SNR has an apparent radius of 5.7\arcmin~and a flux density of 20 to 40 $\upmu$Jy beam$^{-1}$. 

\smallskip
We detected J1914 in OBS2, this time with a significance of 24.3~$\upsigma$, as can be seen in Fig. \ref{fig:pulse}. The results of the detection and analysis to be discussed hereafter are reported in Tab. \ref{tab:fits}. Using the difference in period, $\Delta P$, and time, $\Delta t$, between OBS1 and OBS2 we derive for the first time\footnote{FAST published the timing of one single visit to J1914, based on which a first estimate of the pulsar period was made \citep{Han2021}. The period derivative could not be estimated based on one single observation. } the period derivative, $\dot{P}$ = $\Delta P/ \Delta t$, of J1914 to be 2.7(3)~$\times$~10$^{-14}$~s~s$^{-1}$. Using both the NE2001 \citep{ne2001} and YMW16 \citep{Yao2017} electron density models, we find the distance for a DM of 418.9 pc cm$^{-3}$ to be $d_1$ $\sim$ 7.8 and d$_2$ $\sim$ 8.8 kpc respectively, with the usual caveat of large uncertainties associated with this method. 
This distance range places J1914 (and thus the Mini Mouse) in the same region of the Galaxy hosting the black hole binary GRS 1915+105 (8.6$^{+2.0}_{-1.6}$ kpc, \citealt{Reid2014}). 
At the above distance the Mini Mouse has a projected length between 11.9 and 13.4 pc, and the SNR that we tentatively associated with the pulsar birth has a radius between 12.8 and 14.4 pc (assuming that the remnant is spherically symmetrical). 

\subsection{Localisation}\label{sec:localisation}

Given a flux density of 62.1 $\upmu$Jy and a rms noise of about 50-55 $\upmu$Jy, we do not  expect J1914 to be identifiable as a point source in the image here.
However, we are able to localise J1914 to arcsecond accuracy using S\textsc{ee}KAT,\footnote{\url{https://github.com/BezuidenhoutMC/SeeKAT} by Mechiel C. Bezuidenhout} the Python implementation of TABLo (Tied Array Beam Localiser, \citealt{Bezuidenhout2023}), which is used to localise pulsars discovered by both the TRAPUM and MeerTRAP\footnote{More TRAnsients and Pulsars with MeerKAT} projects (see \citealt{Vleeschower2022, Bezuidenhout2022} for respective examples). S\textsc{ee}KAT implements a novel method of maximum-likelihood estimates using the S/N in each CB and a model of the PSF from \textsc{mosaic}. The PSF changes shape and orientation as the source moves through the sky \citep{Chen2021} so we model it at the observation midpoint. The beams used for the localisation are shown in the left hand panel of Fig. \ref{fig:localisation}. The S/N in each beam was obtained by folding the data with \textsc{dspsr} using an ephemeris of the best period and DM for the observation epoch, before reducing these folded data to a single subintegration and frequency channel and obtaining the S/N using the \texttt{psrstat} command. The results from S\textsc{ee}KAT are shown on the right hand panel of Fig. \ref{fig:localisation}. The position quoted in Tab. \ref{tab:fits} is the OBS2 result because the S/N is higher than OBS1. The weighted mean position of OBS1 and OBS2 is also shown for discussion purposes. 

S/N values in OBS1 disfavour a position within the beam directly on the nebula head, and instead indicate that it lies on its North-Western edge, towards a little spur that may or may not be noise. The analysis of OBS2 yields a bimodal likelihood, seemingly the result of probability leakage towards the gaps in the CB ring\footnote{\textsc{mosaic}'s default is to pack beams within a circle region. Unfortunately, the 6-beam hexagon has gaps due to the ellipticity of the CBs. Care should have been taken when planning OBS2 to make sure the source was observed closer to the zenith, as more circular CBs would better enclose the central beam.}. There is an island of 1 $\upsigma$ contour overlap very closely to the weighted mean position, although both still appear in front of the nebula.

\subsection{Period derivative significance}

As mentioned in \S\ref{sec:results}, we had two MeerKAT epochs that were not phase connected and we used the period and its uncertainty, and standard error propagation to calculate $\dot{P}=$ 2.7(3)~$\times$~10$^{-14}$~s~s$^{-1}$. However, this calculation assumes that the period errors are the dominant source of uncertainty. In order to verify such an assumption,  we calculated other contributions to the uncertainty on $\dot{P}$.
Firstly, we estimated the error in the pulse arrival time at the Earth due to an assumed position error. The R\"{o}mer delay correction for an erroneous source position changes the apparent spin-period as the Earth moves in its orbit \cite[see, e.g.,][]{Lorimer2005,Edwards2006}. 
In the worst case scenario the period is measured at the closest and furthest position from the source on the ecliptic sphere. Using the largest 1~$\upsigma$ error on RA and DEC from Tab. \ref{tab:fits} of 6.2\arcsec~and 13.4\arcsec~and assuming these are approximately equivalent to the uncertainty in ecliptic longitude and latitude, the maximum change in arrival time is 43.4 ms. This yields an apparent change in period of 3.8~$\times$~10$^{-10}$~s over the 6 months, which would translate into a negligible additional $\dot{P}$ of 2.4~$\times$~10$^{-17}$~s~s$^{-1}$. 

Another small apparent change in $P$ could be caused by the Doppler effect from the transverse motion of the pulsar as projected on the celestial sphere, which is known as the Shklovskii effect \citep{Shklovskii1970}. We estimate that the upper limit of this effect may induce an apparent $\dot{P}$ of $\sim$ 7 $\times$ 10$^{-20}$ s s$^{-1}$, which is much smaller than our uncertainty, and thus can be ignored. 

\subsection{X-ray serendipitous observations}
The position of J1914 was observed serendipitously by the Neil Gerhels Swift Observatory (\textit{Swift}) for a total of 8.3 ks, although no source is detected significantly according to the Swift-XRT LSXPS Upper limit server \citep{Evans2023}. We derived an upper limit to the 1-10 keV X-ray flux of approximately 8 $\times$ 10$^{-14}$~erg~cm$^{-2}$~s$^{-1}$, assuming a black-body spectrum with temperature 2~keV and an equivalent column density in the direction of the source of 1.36~$\times$~10$^{22}$~cm$^{-3}$.

\section{Discussion}\label{sec:discussion}

In the MeerKAT observations of the black hole binary GRS 1915+105 we discovered an unusual collimated linear nebula, which based on the similarity with the Mouse feature found in the '80s in the direction of the Galactic Centre \citep{Yusef-Zadeh1987}, we named \textit{the Mini Mouse}. At a distance of 12\arcmin~from the Mini Mouse head, we also identified a circular feature that we classify as a candidate SNR based on its morphology, the apparent centre of which is found on the direction indicated by the tail of the Mini Mouse. 
The relative positions of the Mini Mouse and of the faint candidate remnant are strongly reminiscent of three similar nebulae with radio pulsar engines: the original Mouse \citep{Yusef-Zadeh1987,Camilo2002}, the `Frying Pan' \citep{Kesteven1987, Camilo2009} nebula, and the `Cannon ball' pulsar PSR J0002+6216 \citep{Schinzel2019, Kumar2023}. All three systems have an accompanying SNR, which have been associated with the birth of the escaping pulsars. Such a resemblance motivated our pulsar search. 

Our MeerKAT observations localise J1914 with an accuracy of approximately 10\arcsec~to a position fully consistent with the head of the Mini Mouse, hence supporting a physical association. Thus, we interpret the Mini Mouse nebula as the bow shock produced by the pulsar escaping at supersonic speed through the local ISM away from its birth location. Synchrotron radiation emerges from the pulsar relativistic wind interacting in the nebular magnetic field, similarly to what was concluded for the case of the aforementioned Mouse,  Frying Pan, and cannon-ball nebulae. 

Although the J1914 is localised on the head of the Mini Mouse, its a position lies on the North-Western edge of it rather than to the center of the head. In principle we do not  expect the pulsar to be in front of the radio emission, as the bow shock is formed ahead of the wind termination shock at the boundary where the ram pressure of the ISM equals that of the PWN \citep[e.g.,]{Kargaltsev2017}. However, the ISM likely has pronounced density variability based on the flux variation along the tail in Fig. \ref{fig:cont} so it is possible that the pulsar is moving into a higher density pocket and the the distance between the two shocks is shrinking. An alternative explanation could be that there may be emission on smaller angular scales here that is being missed in the image. Finally, the discrepancy between the continuum map and the pulsar localization inferred from the timing could be explained in terms of a systematic positional offset which may affect MeerKAT images, and is generally of the order 1-2\arcsec~ or larger \citep[e.g., ]{Mauch2020}. High angular resolution observations of the Mini Mouse will clarify this issue.

Our position uncertainty has an area of $7.5~\times~10^{-6}$. The Mini Mouse is directly in the centre of the first Galactic plane quadrant at $l=45^{\circ}$, $b=0^{\circ}$. The region bound by $0^{\circ}<l<90^{\circ}$ and $|b|<~2.5^{\circ}$ contains 910 pulsars, giving a rough probability of chance alignment of 1.5 $\times~10^{-5}$, and hence we are confident that the Mini Mouse nebula is the result of J1914 in motion.

\smallskip

The MeerKAT data yielded a period $P$ and period derivative $\dot{P}$ of 138 ms and $2.7(3)~\times~10~^{-14}$s respectively for J1914, which imply a spin-down luminosity of -$\dot{E}$ = 4$\pi^{2}~I~\dot{P}/P^3$ = $4.0(6)\times10^{35}$~erg~$\mathrm{s}^{-1}$ (where $I$ = 10$^{45}$ g cm$^2$ is the neutron star moment of inertia), a characteristic age $\tau_c = P/2 \dot P = 82$\,kyr, and surface magnetic dipole field strength $B$ =
3.2~$\times$~10$^{19} (P \dot P)^{1/2}$ = 2.1~$\times$~10$^{12}$\,G. These parameters place J1914 near the young and energetic pulsars in the $P$-$\dot{P}$ diagram (see Fig. \ref{fig:PPdot}), characterised by an average spin-down luminosity $\dot{E} \sim 10^{36}$~erg~$\mathrm{s}^{-1}$, and a characteristic age $\tau_c$ ranging from 10 to 100 kyr. Given the upper limit to the X-ray luminosity of the Mini Mouse, assuming isotropic emission, we obtain an upper limit to the X-ray luminosity of L$_{\rm X} < 7 \times 10^{32}$erg/s, and thus an upper limit to conversion efficiency of the spin-down luminosity into X-ray emission of L$_x$/$\dot{E} <$ 1.5 $\times $ 10$^{-3}$. The non-detection is not surprising as this limit is a factor of about 6 higher than the limit estimated for the `Frying Pan pulsar', which is also undetected in X-rays \cite{Camilo2009}. 

If we assume that the pulsar actual age is approximately equal to its characteristic age (which is not necessarily true, see \citealt{Gaensler2000}), the SNR should be approximately 80~kyr old. At such an age a typical remnant in a low density ambient medium is expected to have entered the Snowplough expansion phase, having transitioned from the Sedov-Taylor adiabatic expansion phase about 50 kyr after detonation \cite[see][and references therein]{Jimenez2019}. In the Snowplough phase the radio shell should have a very low surface brightness $\Sigma$, and large diameter $D$ according to the $\Sigma-D$ inverse relationship \cite[see, e.g.,][]{Vukotic2019}. At our DM-based distances the SNR shell would have a diameter of 12.8 pc or 14.4 pc, which is more typical of a remnant in the mid-to-late Sedov-Taylor phase, with an age of about 20 kyr. Such a discrepancy can be explained if the ISM is sufficiently dense \citep{Frail1994}, as efficient radiative cooling from expansion into a high density ambient medium can inhibit the Sedov-Taylor phase, a plausible scenario given that this remnant is located on the Galactic plane \cite[see, e.g.][]{Terlevich1992,Jimenez2019}. We note that the distance derived from the DM measure might not be accurate, and hence we could in principle decide to ignore it and instead derive the distance from the $\Sigma-D$ relation. However, this method has often been treated with scepticism due to possible intrinsic biases in the correlation, and in particular due to the fact that (i) the $\Sigma-D$ relation is derived from independently determined distances, which each have their own uncertainties \citep{Green1991}; (ii) SNRs show a diversity of intrinsic properties, for example the population of SNRs in the LMC that have are roughly all the same distance from Earth \citep{Green1984}. Therefore we decided to wait for additional information on the source to make this attempt.

Again assuming that the pulsar characteristic age of 82 kyr is close to the pulsar actual age, and that the faint circle of emission is the progenitor remnant, we can obtain an estimate of the pulsar's transverse velocity component, $v_{\perp}$, by considering the distance between the centre of the SNR and the head of the Mini Mouse. For a pulsar distance of 7.8 or 8.8 kpc the resulting projected pulsar velocity is between 320 and 360 km/s, which is well within the kick velocity distribution for young, isolated pulsars, centred at approximately 300 km/s, with a dispersion of approximately 190 km/s \citep{Hansen1997}.  

If the connection between J1914 and the faint SNR is correct, then we may have a faint, fast-spinning, distant young pulsar with a high kick velocity, i.e. a member of an under-sampled population, which could help extrapolating the local young pulsar velocity distribution to the wider Galactic one \citep{Hansen1997}.
Under the assumption that the magnetic moment of the pulsar remains constant, the actual age and the characteristic age of a pulsar are connected by the relation $\tau = \tau_{\text{c}}[1 - P_0/P)^{n-1}]/(n -1) $, where $P_0$ is the pulsar spin period at birth and $n$ is the breaking index of rotation \citep{Manchester1978}. Using the standard assumptions of $P_0$~$\ll$~$P$ and $n$ = 3 in the case of magnetic dipole breaking \citep[see, e.g.][for the case of the Mouse pulsar]{Camilo2002} and $\tau = \tau_{\text{c}}$, the above relation implies that P$_0$ is between 10-20 ms, which is consistent with the expected theoretical values for a newborn radio pulsar and thus supports our hypothesis of a young and fast spinning pulsar.  

\section{Summary and conclusions}\label{sec:conclusions}

In the MeerKAT data dedicated to the monitoring of a Galactic X-ray binary, we found a radio nebula, which we named "the Mini Mouse". Such a feature is produced by a supersonic pulsar escaping the location of its birth, marked by the presence of the faint and previously unknown supernova remnant candidate G45.24$+$0.18 also discovered in the MeerKAT data. 
Time-domain observations found the signal from a radio pulsar in correspondence of the tip of the Mini Mouse, confirming the association of the nebula with a previously poorly localised faint pulsar discovered by the FAST telescope, J1914+1054g. 

The Mini Mouse is the fourth case of a bow shock associated with an escaping pulsar, for which both the pulsar signal and the SNR associated with its birth have been observed. Additional high angular resolution observations of the Mini Mouse will clarify its finer scale structure, and will bring a better characterisation of the local ISM, while additional high time-resolution observations of J1914+1054g will significantly improve the timing solution, thus providing a better constraints on the pulsar properties. 

The ThunderKAT discovery of the Mini Mouse and the TRAPUM confirmation of the association with J1914+1054g constitute an excellent example of immense potential of the MeerKAT data. Thanks to the detection of structures similar to the Mouse and Mini Mouse, MeerKAT will help unveiling more young radio pulsars which will add to the still small population of such objects, which is predicted to count thousands of members in our Galaxy \citep[e.g.][]{Lorimer1993}.

\section*{Acknowledgements}

The authors thank the anonimous referee that reviewed this work providing useful suggestions to improve it. 
SEM acknowledges JinLin Han for providing the spin period of PSR J1914+1054g from the FAST GPPS, and Lauren Rhode for comments on an early version of this work.
SEM acknowledge financial contribution from the agreement ASI-INAF n.2017-14-H.0 and from the INAF mainstream grant. JDT acknowledges funding from the United Kingdom's Research and Innovation Science and Technology Facilities Council.
All the authors thank the staff at the South African Radio Astronomy Observatory (SARAO) for scheduling the MeerKAT observations presented here. TRAPUM observations used the FBFUSE and APSUSE computing clusters for data acquisition, storage and analysis. These clusters were funded and installed by the Max-Planck Instit{\"u}t f{\"u}r Radioastronomie (MPIfR) and Max-Planck-Gesellschaft. The MeerKAT telescope is operated by the South African Radio Astronomy Observatory, which is a facility of the National Research Foundation, an agency of the Department of Science and Innovation.
SAOImage DS9 was used for image analysis and presentation.

\section*{Data Availability}

The un-calibrated MeerKAT visibility data presented in this paper are publicly available in the archive of the South African Radio Astronomy Observatory at https://archive.sarao.ac.za. 
The Continuum MeerKAT observations were taken as part of the ThunderKAT Large Survey Program, project code SCI-20180516-PW-01. 
The project code for the TRAPUM Science Working Group collaborated with is SCI-20180923-MK-03, and for the DDT observation is DDT-20221028-SM-01.
Data that are not available thorough public archives, and all source code, will be shared on reasonable request to the corresponding author.



\bibliographystyle{mnras.bst}
\bibliography{biblio}








\bsp	
\label{lastpage}
\end{document}